\documentclass[12pt]{article} 

\usepackage{ol2}

\begin{document}


\title{The relation between optical beam
propagation in free space and in strongly nonlocal nonlinear media}
%

\author{Daquan Lu, Wei Hu$^*$, and Qi Guo}
\address{
Laboratory of Photonic Information Technology, South
China Normal University, Guangzhou 510631, China\\
$^*$huwei@scnu.edu.cn }

\begin{abstract}
The relation between  {optical beam} propagation in
strongly nonlocal nonlinear (SNN) media and  {propagation} in free
space is  {demonstrated using} the technique of variable
transformation. The governing equation, integral and analytical
solutions, and propagation properties in free space can be directly
 {transferred} to  {their counterparts} in SNN media
through a one-to-one correspondence. The one-to-one correspondence
together with the Huygens-Fresnel integral yields an efficient
numerical method  {to describe} SNN propagation.  {The existence
conditions and possible structures of solitons and breathers in SNN
media are described in a unified manner by comparing propagation
properties in SNN media with those in free space.}
 The results can be employed in other contexts in which the
governing  {equation for the evolution of waves is} equivalent to
that in SNN media, such as  {for} quadratic graded-index media,
 {or for} harmonically trapped Bose-Einstein condensates in
the noninteracting limit.
\end{abstract}
\ocis{190.6135, 190.4420, 190.5940, 260.1960.}
\maketitle 

\section{Introduction}

 The propagation properties of light beams in nonlocal
nonlinear media have attracted   {significant} attention in recent
years. There are {various interesting} properties induced by the
nonlocality, such as the suppression of   collapse \cite{bang02}
 {and} the support of vortex solitons \cite{Yakimenko06}
and multi-pole solitons \cite{Buccoliero07}.
 {For} the special case of the strongly nonlocal nonlinear
(SNN) media in which the characteristic length of the material
response function is much larger than the beam width, the
propagation equation can be linearized to the well-known
Snyder-Mitchell model (SMM) \cite{Snyder97}. In fact, since Snyder
and Mitchell introduced the SMM to investigate  {optical beam}
propagation in SNN media, various soliton solutions
\cite{Yakimenko06,Buccoliero07,
Krolikowski00,Deng07,Mamaev97,Nikolov04,Aguayo07,ouyang}, such as
Hermite-Gaussian (HG) \cite{Buccoliero07}, Laguerre-Gaussian (LG)
\cite{Buccoliero07} and Ince-Gaussian (IG) \cite{Deng07,Aguayo07}
solitons, have been  {predicted theoretically.} Some soliton
structures and their interaction have been observed experimentally
in SNN materials such as nematic liquid crystal
\cite{Conti04,Peccianti02b}, and lead glass \cite{Rotschild06}.

 {In addition}, to the best of our knowledge, the propagation in
free space has been investigated more thoroughly than the other
propagation problems in the field of optics. The paraxial
diffraction equation
 has been  {thorughly} investigated.
Various types of  beam solutions with different transverse profiles
have been obtained in Cartesian, circular cylindrical, and
elliptical coordinates (e.g., \cite{Bandres07,Bandres08,Bandres06}
and references therein). These solutions can be roughly classified
into two types: (i) shape-invariant beams, such as LG, HG, and IG
beams;
 and (ii) shape-variant beams, such as higher order
elegant-Hermite-Gaussian (EHG), elegant-Laguerre-Gaussian (ELG), and
elegant-Ince-Gaussian (EIG) beams. The propagation of these beams
has been  {investigated in detail and} many parameters, such as
width, divergence,  {radius of curvature,}  and quality factor, have
been introduced to describe  {their propagation}. In {summary}, the
theory of free propagation has been  {well developed over}  the past
decades.

Compared with that of the free propagation, the problem of SNN
propagation  is mathematically much  {more} complicated \cite{guo},
even for the problem of soliton which  {may} be the simplest
 {example} of the propagation problems in SNN media.
However, the structures of the HG, LG and IG solitons introduced in
the literature are also the modes in free space \cite{lasers},
 {leading naturally to the question of whether} any direct
relation exists between   { free propagation and propagation} in SNN
media {.}

In fact, if there exists a one-to-one correspondence between
 {free propagation and propagation in SNN media},
 {the results of free propagation theory could be applied to
the following aspects of the study of SNN propagation}: (i) The
structures  {of} and the existence conditions  {for} breathers and
solitons  {could} be conveniently described with simple and
intuitional physical pictures of free propagation in a unified
manner. (ii)  {Whereas} the previous investigations are mainly
focused on solitons and breathers  {that} are shape-invariant upon
propagation, the shape-variant propagation in SNN media remains
unexplored. With  {an established correspondence between free and
SNN propagation}, it would be easy to deal with the propagation of
an arbitrary  field in SNN media  { by using the well developed}
theory of free propagation, avoiding complicated mathematical
 {calculations}. (iii) The well developed parameters of free
propagation  {could} be directly transplanted to  {the} SNN case to
 {characterize} the propagation  {properties}.
 (iv) In experiments,  beams are usually transmitted from free
 space into SNN media. A direct  {correspondence} between
   {free propagation  and propagation in SNN media} would be of
 practical interest for designing of experiments.

In this letter, we  {describe our research into} the relation
between the SNN propagation  and the free propagation, and describe
the propagation in SNN media  {using the well-developed} theory and
 {clear} physical pictures of free propagation. The relation
between  propagation in SNN media and that in free space is
 {studied using the variable transformation technique}.
 {This} shows that the governing equation, beam solutions,
and propagation properties  {for free propagation} can be directly
transplanted  {with a one-to-one correspondence} to  {the case of
propagation} in SNN media. On the basis of the one-to-one
correspondence and the Huygens-Fresnel integral,  {we develop} an
efficient numerical method  {to describe}  SNN propagation.
 {We describe the existence conditions and possible
structures of solitons and breathers in SNN media in a unified
manner by comparing the propagation properties in SNN media with
those in free space. As an example, the theoretical predictions are
illustrated for the case of EHG beams.}

\section{ One-to-one correspondence}

 {We begin by using the technique of variable transformation
to connect propagation in SNN media to free propagation.}
 This technique is
frequently used  {to study} similaritons in nonlinear wave guides
(e.g., \cite{Kruglov03,Ponomarenko07}).  { The variable
transformation technique allows} the governing equation [e.g., the
inhomogeneous nonlinear Schr\"{o}dinger equation (NLSE)]
 {to} be reduced to a   {simpler mathematical form}
(e.g., the standard NLSE), and the solution of the former can be
obtained from  {the solution} of the latter  {by exploiting} a
one-to-one correspondence. Because the structures of the HG, LG and
IG solitons in SNN media are also the modes in free space
\cite{lasers},  { it is reasonable to expect that  a transformation
exists that} can reduce the governing equation of SNN propagation to
that of free propagation.

In  {the} laboratory reference frame,  {beam propagation } in
nonlocal nonlinear media is governed by the nonlocal NLSE
\begin{equation}
 2ikn_0{\partial_z A} + n_0 (\partial_{xx}+\partial_{yy}) A +
2k^2{\triangle n} A=0,\label{1}
\end{equation}
where $k$ represents the wave number in the media with a linear
refractive index $n_0$,  $\triangle n = n_2 \int {R(\mathbf r -
\mathbf r_a)} |\Phi|^2 {\rm{d}}^2 \mathbf {r}_a$ represents the
nonlinear perturbation of refractive index, where $n_2$ is the
nonlinear index coefficient, $R$ is the normalized symmetric real
spatial response function of the media,  {and} $\mathbf {r}= (x, y)$
 {represents the 2-dimensional transverse coordinate
vector.}

In the case of SNN media the nonlocal NLSE can be simplified to
 {the} modified SMM \cite{ouyang}
\begin{equation}
2ik{\partial_{z^\prime}  \Phi } + (\partial_{x^\prime
x^\prime}+\partial_{y^\prime y^\prime}) \Phi - k^2 \gamma^2 P_0
r^{\prime 2} \Phi = 0\label{2}
\end{equation}
in a new reference frame  {that} moves with  {the center of mass,}
\begin{equation}
z^\prime=z, \ \ \ \mathbf r^{\prime}=\mathbf r-\mathbf r_c(z).
\end{equation}
The adoption of this reference frame is important for the input
fields whose transverse spatial momentum is  {nonzero}
\cite{ouyang}. In this reference frame the field becomes
\begin{equation}
\Phi (\mathbf r^\prime, z^\prime)=A ( \mathbf r^\prime+\mathbf r_c ,
z^\prime) \exp [-\frac{ik\mathbf M\cdot (\mathbf r^\prime+\mathbf
r_c)}{P_0}+\frac{ikM^2z^{\prime}}{2P_0^2}].\label{basic transform}
\end{equation}
In Eqs. (\ref{2})-(\ref{basic transform}), $\gamma$ is a material
constant, $P_0 = \int |\Phi|^2 {\rm{d^2}}\mathbf r ^{\prime}$ is the
input power, $\mathbf r_c (z^{\prime}) = \mathbf r_c (0) + \mathbf
Mz^{\prime}/P_0$ is  {the center of mass of} the beam, $\mathbf M =
(i/2k)\int {(A\nabla _ \bot A^*  - A^* \nabla _ \bot A)}
\rm{d}x\rm{d}y$ is the  transverse spatial momentum,   $\mathbf
{r^{\prime}}= (x^{\prime}, y^{\prime})$,  {and} $\mathbf {r_c}=
(x_c, y_c)$.

To connect  the governing equation of SNN  {propagation} with that
of free propagation, we adopt the transformations
\begin{equation}
\left\{
\begin{array}{l}
 \mathbf{r}^{\prime}=(-1)^{a}\frac{w_{c0}}{w_c(\zeta)}{\mathbf{s}}\\
 {z}^{\prime}=z_{c0}[\arctan (\frac{\zeta}{z_{c0}})+a\pi]\\
 \Phi(\mathbf{r}^{\prime},{z}^{\prime})=(-1)^{a}\frac{w_c(\zeta)}{w_{c0}}
 \exp [-\frac{ik{\mathbf{s}}^2}{2R_c(\zeta)}]\Psi ({\mathbf{s}},\zeta)
 \end{array} ,\right.\label{transform}
\end{equation}
where $w_c(\zeta)={w_{c0}}[1+({\zeta}/{z_{c0}})^2]^{{1}/{2}}$,
$R_c(\zeta)=\zeta[1+({z_{c0}}/{\zeta})^2]$, $z_{c0}=kw_{c0}^2$,
$w_{c0}= ({{k^2 \gamma^2 P_0 }})^{-{1}/{4}}$,
$a=0,1,-1,2,-2,\cdot\cdot\cdot$,  {and} ${\mathbf{s}}= (\mu, \nu)$.
Then Eq. (\ref{2})  {reduces} to
\begin{equation}
 (\partial_{\mu\mu}+\partial_{\nu\nu}) \Psi+2ik{\partial_ \zeta}
\Psi  = 0. \label{freespace}
\end{equation}
Equation (\ref{freespace}) is  the well-known paraxial diffraction
equation  {that} governs the paraxial  propagation of monochromatic
 {beams} in free space. Thus,  {Eqs.
(\ref{2})-(\ref{freespace}) establish} the one-to-one correspondence
between the beam solution in SNN media and that in free space:
\begin{equation}
\Phi(\mathbf{r^{\prime}},z^{\prime})=F_1F_2\times
\Psi({F_1}\mathbf{r^{\prime}},F_3),
 \label{free2nonlocal}
\end{equation}
where
\begin{equation}
\left\{
\begin{array}{l}
 F_1(z^{\prime})=(-1)^{a}[1+\tan^2(\frac{z^{\prime}}{z_{c0}})]^{\frac{1}{2}}\\
 F_2(\mathbf{r^{\prime}},z^{\prime})=\exp\{-\frac{ikF_1(z^{\prime})^2{r^{\prime 2}}}
 {2z_{c0}[\tan(\frac{z^{\prime}}{z_{c0}})+{1}/{\tan(\frac{z^{\prime}}{z_{c0}})}]}\}
 \\
 F_3(z^{\prime})=z_{c0}\tan(\frac{z^{\prime}}{z_{c0}})
\\a(z^{\prime})=\frac{1}{\pi}\{\frac{z^{\prime}}{z_{c0}}-\arctan[\tan(\frac{z^{\prime}}{z_{c0}})]\}
\\z_{c0}(P_0)=\frac{1}{\sqrt{P_0}\gamma}
 \end{array} \right..\label{transform2}
\end{equation}
 {Note} that $z_{c0}$ is not a constant, but varies with the
input power $P_0$. Equation (\ref{free2nonlocal}) connects the
propagation in SNN media
 with  {free propagation}.
  The  {numerous free space monochromatic
   beam solutions and propagation properties}
 can be conveniently transplanted to  {their counterparts} in SNN media
 {using} Eq. (\ref{free2nonlocal}).

Since  { a one-to-one correspondence exists between the beam
solutions in SNN media and those in free space}, a general
comparison between the propagation  {properties} in SNN media and
 {those} in free space would be constructive.  {We
therefore compare  three beam propagation properties.}

  {Beginning with beam patterns, we see from Eq. (\ref{free2nonlocal}) that}
the beam in SNN media evolves periodically with   {a} period
$\triangle z=2\pi{z_{c0}}$. For convenience of discussion, we divide
each period (from $z^{\prime}=(2a-1/2)\pi z_{c0}$ to
$z^{\prime}=(2a+3/2)\pi z_{c0}$) into two half-period. In the
 {leading} half-period  {(from
$z^{\prime}=(2a-1/2)\pi z_{c0}$ to $z^{\prime}=(2a+1/2)\pi
z_{c0}$)}, the evolution of the pattern is a condensed configuration
of that in free space from $-\infty$ to $+\infty$. There is a
 {one-to-one correspondence} between patterns in SNN media
and those in free space, i.e., the pattern shape at the cross
section $z^{\prime}$ in SNN media is the same as that at the cross
section $\zeta=z_{c0}\tan(z^{\prime}/z_{c0})$ in free space. At
special cross sections where $z^{\prime}/z_{c0}-2a\pi=-\pi/2$,
$-\pi/4$, $0$, $\pi/4$, $\pi/2$, the  {beam pattern shapes} in SNN
media are respectively the same as that at $\zeta=-\infty$, $
-z_{c0}$, $ 0$, $z_{c0}$, $+\infty$ in free space. In the
 {trailing} half-period  {(from
$z^{\prime}=(2a+1/2)\pi z_{c0}$ to $z^{\prime}=(2a+3/2)\pi
z_{c0}$)}, the beam patterns  {are} the reverse of those in the
 {leading} half-period, and the evolution of the pattern is
corresponding to that of a inverse field (
$\Psi(-{\mathbf{s}},\zeta)$) in free space. In fact, patterns of
most beams are symmetrical. For these beams, the beam patterns in
the  {trailing} half-period  {are} the same as those in the
 {leading} half-period, therefore the period of pattern
evolution  {reduces} to $\triangle z=\pi{z_{c0}}$.

 {The second beam property we compare is the beam width.}
Although the pattern shape at the cross section $z^{\prime}$ in SNN
media is the same as that at the cross section
$\zeta=z_{c0}\tan(z^{\prime}/z_{c0})$ in free space, the beam width
in SNN media  {decreases} by a factor of $|F_1(z^{\prime})|$
compared  {with} that in free space.  {Explicitly},
\begin{equation}
w^{(s)}(z^{\prime})|_{z^{\prime}=z^{\prime}}
=\frac{w^{(f)}(\zeta)|_{\zeta=z_{c0}\tan(z^{\prime}/z_{c0})}}{|F_1(z^{\prime})|}
\label{wfs}
\end{equation}
 {where the superscripts (f) and (s) refer to the cases of
propagation in free space and in SNN media, respectively, and
$w^{(s)}$ and $w^{(f)}$  are the corresponding beam widths. Equation
(\ref{wfs}) results} from
 the self-focusing effect of SNN media.   {Accordingly} the amplitude
  {increases}  by a factor of $|F_1(z^{\prime})|$,
  {in accordance with energy conservation.}

 { The final propagation property we compare involves the
cophasal surfaces. } For convenience of discussion, we assume the
radius of the cophasal surface at the cross section
$\zeta=z_{c0}\tan(z^{\prime}/z_{c0})$ in free space is $R^{(f)}
(\mathbf s, \zeta)$, so that in free space the phase variation
across the transverse plane can be written as
$\exp[ikr^2/2R^{(f)}]$. In SNN media, according  to Eq.
(\ref{free2nonlocal}), the phase variation across the transverse
plane at the corresponding cross section $z^{\prime}=z^{\prime}$
would be $\exp[ikr^2/2R^{(s)}(z^{\prime})]$, where
$R^{(s)}(z^{\prime})$ is the radius of the cophasal surface in SNN
media,  {and is given by}
\begin{equation}
R^{(s)}(z^{\prime})=\frac{1}{\frac{F_1(z^{\prime})}{R^{(f)}(\mathbf
s, \zeta)}-\frac{F_1(z^{\prime})}{R_c(\zeta)}}.
\end{equation}
In  {the} special case  {where} $R^{(f)}(\mathbf s,
\zeta)=R_c(\zeta)$, $R^{(s)}(z^{\prime})$ approaches infinity, the
 {cophasal} surface   {remains} planar
 {upon} propagation.

 {The evolution of the beam width and  of the cophasal
surfaces is interdependent for SNN propagation.} For example, when a
HG beam is input at the waist and the relation $z_R=z_{c0}$ (where
$z_R$ is the Rayleigh distance) is satisfied, the free propagation
 increases the beam width by a factor of $F_1(z^{\prime})$
(i.e., $w^{(f)}(\zeta)= F_1(z^{\prime})w^{(f)}(0)$), so that in SNN
media the beam width   {remains} invariant during propagation (i.e.,
$w^{(s)}(z^{\prime})=w^{(f)}/F_1(z^{\prime})=w^{(f)}(0)$).
Simultaneously, the  {cophasal} surface    {remains} planar during
propagation, because $R^{(f)}(\mathbf s, \zeta)=R_c(\zeta)$ is
satisfied.
  {If the beam is not input at the waist and/or
  the relation $z_R=z_{c0}$ is not satisfied,
the beam width and the cophasal surface} both  evolve periodically
 {during} propagation. This property is important for the
existence of solitons and breathers, as will be discussed
 {below}.

\section{Numerical   {description} of SNN propagation}

In real optical systems, there exist many beams with irregular
amplitude and phase profiles.   {Numerical methods play} an
important role in   {studying} these situations. The most
 {popular} numerical method is the split-step Fourier method
(SSFM).  The  SSFM is   {based on
 dividing the propagation length}
 into a large number of segments  {and assuming that}
in each segment the diffractive and nonlinear effects  {are
independent and may be}  calculated separately. Although the SSFM is
much faster than the finite-difference method, it is still
time-consuming, because of the large number of segments which is
required to ensure the accuracy. Since in  {the} SNN case the
nonlocal NLSE can be simplified to the SMM model  {( connected to
free propagation)}, we develop a simple numerical method
 {that is more efficient than the SSFM.}

 In the case of free propagation, an efficient numerical
method  {exists} based on the integral solution of Eq.
({\ref{freespace}}), i.e., the  Huygens-Fresnel integral
\cite{lasers}
\begin{eqnarray}
 \Psi( {\mathbf{s}},\zeta)=\frac{-ik}{2\pi \zeta}\int
 \Psi( {\mathbf{s}}_0,0)\exp[\frac{ik}{2\zeta}|
  {\mathbf{s}}- {\mathbf{s}}_0|^2]\rm{d^2\mathbf {\mathbf{s}}_0}.\label{fresnel}
\end{eqnarray}
Because Eq. (\ref{fresnel})  {represents a} convolution of the input
field with a spherical  {wave function}, the field at any plane can
be obtained easily from the input plane by using the fast Fourier
transform algorithm.

Since there is a one-to-one correspondence between the beam solution
in SNN media and that in free space, we  get the integral solution
in  SNN media using Eqs. (\ref{free2nonlocal}) and (\ref{fresnel}),
 {as follows}:
\begin{eqnarray}
\Phi(\mathbf{r^{\prime}},z^{\prime})= \int
\varphi(\mathbf{r}^{\prime}, \mathbf{r}^{\prime }_0)
 \Phi(\mathbf r^{\prime}_0,0){\rm{d^2}}\mathbf r^{\prime}_0,\label{frft}
\end{eqnarray}
where
\begin{eqnarray}
 \varphi(\mathbf{r}^{\prime}, \mathbf{r}^{\prime
}_0)&=& \frac{-i}{2\pi w_c^2\sin
(\frac{z^{\prime}}{z_{c0}})}\nonumber
\\
&\times&\exp [\frac{{ir^{\prime 2}+ ir^{\prime 2}_0-2i
 \mathbf{r}^{\prime}\cdot \mathbf{r}^{\prime }_0\sec(\frac{z^{\prime}}{z_{c0}})  }}{{2 w_c^2\tan
(\frac{z^{\prime}}{z_{c0}}) }} ].
\end{eqnarray}
 Equation (\ref{frft}) is equivalent to  $
\Phi(\mathbf{r^{\prime}},z^{\prime})=\hat{F_\alpha  }\{ \Phi(\mathbf
r^{\prime}_0,0)\} e^{ - i\alpha }$, where $\hat{F_\alpha}$
represents the fractional Fourier transform  {of} order
$\alpha={z^{\prime}}/{z_{c0}}$. For this reason,  {in a separate
publication\cite{lu} we refer to } the propagation in SNN media
 {as} the self-induced fractional Fourier transform.

 The numerical
simulation of the SNN propagation in the laboratory reference frame
can  {therefore} be accomplished in three steps:  {First}, we
calculate the initial  {center of mass} $\mathbf r_c(0)$ as well as
the transverse momentum $\mathbf M$  {and use Eq. (\ref{basic
transform}) to} transform the field in the laboratory reference
frame (i.e. $A ( \mathbf r, z)$) to  {the field} in the reference
frame $ z^\prime=z, \mathbf r^{\prime}=\mathbf r-\mathbf r_c$ (i.e.
$\Phi (\mathbf r^\prime, z^\prime)$). { Next, we propagate the field
in the reference frame $ z^\prime=z, \mathbf r^{\prime}=\mathbf
r-\mathbf r_c$ from the input plane $ z^\prime=0$ to the later plane
$ z^\prime=z^\prime$ using Eq. (\ref{frft}).}  {Finally, we}
transform the field at the plane $ z^\prime=z^\prime$ in the
reference frame $ z^\prime=z, \mathbf r^{\prime}=\mathbf r-\mathbf
r_c$ to that in the laboratory reference frame  {using} the inverse
transformation
\begin{equation}
A ( \mathbf r, z)=\Phi (\mathbf r-\mathbf r_c, z) \exp
[\frac{ik\mathbf M\cdot \mathbf r}{P_0}-\frac{ikM^2z}{2P_0^2}].
\end{equation}

This approach provides a  straightforward way to numerically
propagate any field from the input plane to an arbitrary later plane
in  SNN media. Since the propagation distance is not required to be
divided into a large number of segments, only  {a single} fractional
Fourier transform is required,  {making this approach} much more
efficient than the SSFM  {for the} SNN case. We believe the fast
fractional Fourier transform algorithm which has been developed in
recent years (e.g. \cite{yang04} and references therein) would make
this approach  {even} more efficient.

\section{Breathers and  solitons
in SNN media}

A special feature of SNN media is that the nonlocality  {
 supports (2+1)D solitons and breathers
  \cite{bang02,Buccoliero07,Deng07,Aguayo07},
  preventing  the catastrophic  collapse  that occur in local nonlinear media.} Here,
   {using} Eq. (\ref{free2nonlocal})
and  {comparing}  propagation  {properties} in SNN media to those in
free space, the existence conditions of
  breathers and  solitons in SNN media can be
 {described conveniently} in a unified manner.  {If
a beam keeps its shape during free propagation, it would be a
breather in SNN media, because the beam shape would remain invariant
and the beam width as well as the cophasal surface would evolve
periodically with the period  $\triangle z=\pi z_{c0}$ in SNN
media.} Furthermore, if the input power  {and} the entrance plane
 {are} designed appropriately so that the beam width and the
beam shape simultaneously  {remain} invariant  {during} SNN
propagation, the breather would reduce to a soliton.

This explains why the HG, LG, and IG  solitons exist in the SNN
media.  {They retain} the beam shape in free space,
 {therefore} they generally evolve as breathers in SNN
media. In the special case that the field is input at the beam waist
and the input power $P_0$  {equals} the critical power $P_c$
 { (so that $z_{c0}=z_R$, where $P_c=1/(k^2\gamma^2
w_{0}^4)$,}  and $z_R$ is the Rayleigh distance of the input field),
 {then  diffraction increases and self-focusing decreases}
the beam width by the same factor of $F_1(z^{\prime})$, and the
deforming of the cophasal surfaces caused by  self-focusing exactly
balances that caused by  diffraction. Therefore  the beam width in
addition to the beam shape  {remains} invariant, and the breather
 {reduces} to a soliton.

 {Furthermore}, based on  {the analysis above}  we
can extend the range of breathers and solitons in SNN media to the
input fields  {that} are linear  {superpositions} of the degenerate
solutions of HG ,LG, and IG beams with the same Rayleigh distance,
 {beam waist location}, and Gouy phase shift in free space.
These superposed fields are shape-invariant in free propagation,
thus  {they propagate } as solitons in SNN media when  {the entrance
plane is located at the beam waist} and $P_0=P_c$, otherwise
 {they propagate } as  {breathers}. The various
structures of these superposed fields would greatly enrich the
family of solitons and breathers.

\section{Example}

To illustrate  { the predictive capacity of our calculations}, we
take the EHG beams as an example. In free space, the field of the
$(m, n)$ mode EHG beam can be written as \cite{lasers}
\begin{eqnarray}
\Psi_{mn}
({\mathbf{s}},\zeta)=\psi[\frac{q_0}{q(\zeta)}]^{\frac{m+n}{2}+1}
H_m(\sqrt {c(\zeta )} \mu)\nonumber
\\
\times\ H_n(\sqrt {c(\zeta )} \nu)\exp[-c(\zeta )s^2 ],
\end{eqnarray}
where $c(\zeta)=-{ik}/{2q(\zeta)}$, $q(\zeta)=\zeta-iz_R$,
$z_R=kw_0^{2}$ is the Rayleigh distance,  {and} the coefficient
$\psi$ is determined by the input power through $P_0 = \int |\Psi|^2
{\rm{d^2}}\mathbf{s}$. In free propagation, the $(m, n)$ mode EHG
beam is shape-invariant when $m, n \leq  1$, otherwise it is
shape-variant.

Obtaining the field of EHG beams in SNN media  {using SMM directly
is } mathematically complicated.  {However, using} Eq.
(\ref{free2nonlocal}) it  {is} easily obtained as:
\begin{eqnarray}
\Phi_{mn} ({\mathbf{r^\prime}},z^\prime)&=&
F_1(z^\prime)F_2({\mathbf{r^\prime}},z^\prime)\psi[\frac{q_0}{Q(z^\prime)}]^{\frac{m+n}{2}+1}
\nonumber
\\
&\times& H_m[\sqrt {C(z^\prime)} F_1(z^\prime)x^\prime]
  H_n[\sqrt {C(z^\prime)} F_1(z^\prime)y^\prime]\nonumber
\\
&\times&\exp\{-C(z^\prime)[F_1(z^\prime)r^{\prime }]^2\},
\end{eqnarray}
where $C(z^\prime)=-{ik}/{2Q(z^\prime)}$,
$Q(z^\prime)=z_{c0}\tan({z^\prime}/{z_{c0}})-iz_{R}$,
$z_{c0}={1}/{\gamma\sqrt {p_0} }$.

\begin{figure}[t]
  \centering
  \includegraphics[width=10cm]{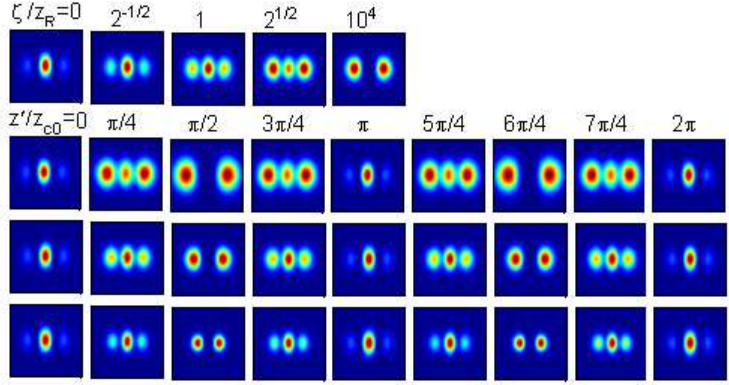}
  \caption{ (Colour online) Evolution of the pattern  of the (2, 0) mode EHG beam in free
  space (row 1) and in SNN media (rows 2-4). The input power of the
  SNN propagation is $P_0=0.5P_c$ (row 2), $P_c$ (row 3), and $2P_c$ (row 4),
  respectively, where $P_c=1/(k^2\gamma^2 w_{0}^4)$.
  In row 1 the transverse dimension is scaled by a factor of
  $1/[1+(\zeta/z_R)^2]^{1/2}$,  whereas in other rows it is not scaled.}
  \end{figure}
\begin{figure}[t]
  \centering
  \includegraphics[width=12cm]{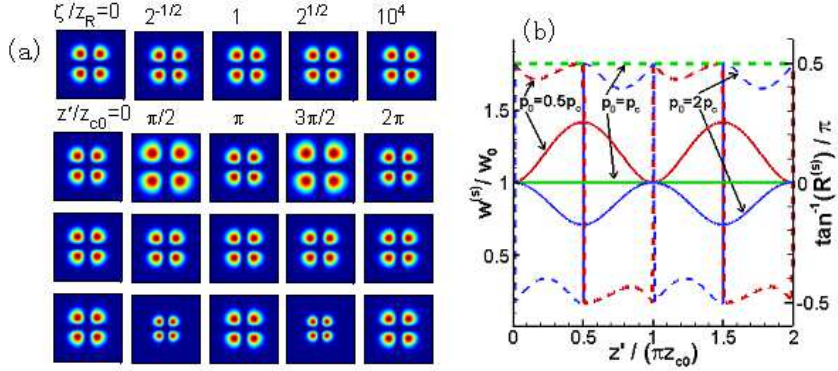}
  \caption
  { (Colour online) (a) The same as Fig. 1 except that the input field is a (1,1) mode EHG beam.
  (b) Evolution of the beam width (solid lines) and the radius
  of cophasal surfaces (dashed lines).
 }
  \end{figure}

Figure 1  {shows} the evolution of the pattern of the (2, 0) mode
EHG beam in  {a SNN medium under varying  input power compared with
the same situation for free propagation}.  {We find that  if the
pattern varies upon propagation in free space, it does so in the SNN
medium as well.}
 Because the transverse pattern is
distributed symmetrically, the pattern evolves periodically with the
period ${\triangle z^\prime}=\pi{z_{c0}}$.  {Furthermore}, at the
cross  {section} $z^\prime=(a+1/2)\pi z_{c0}$, the pattern is the
same as that at the far field in free space  {and does} not vary
with the input power,  {and the beam width $w$ is inversely
proportional to $\sqrt{P_0}$, because the field at the cross section
may be regarded as the conventional Fourier transform of the input
field.} At an arbitrary cross section where $z^\prime\neq a\pi
z_{c0}/2$, the pattern shape is the same as that at the
 {free space} cross section
$\zeta=z_{c0}\tan(z^{\prime}/z_{c0})$. Since $\zeta$ varies with the
input power, the pattern is different for different input power. For
example, at $z^\prime=\pi z_{c0}/4$, the pattern for the input power
$P_0=0.5P_c$, $P_c$, $2P_c$ is the same as  {the free space pattern}
at $\zeta=\sqrt{2} z_R$, $z_R$, $z_R/\sqrt{2}$, respectively.

Because the   ($1, 1$) mode EHG beam is shape-invariant in free
space, its pattern  {remains} invariant in SNN media (Fig. 2).
Generally,  { the  ($1, 1$) mode EHG in SNN media propagates as a
breather}, i.e., the beam width as well as the radius of
 {the} cophasal surfaces varies periodically with the period
$ \triangle z=\pi z_{c0}$.
 At special cross sections where $z^\prime/z_{c0}=a\pi/2$
  { the width is either
 maximum or minimum (in fact, the width reaches its maximum and
 minimum alternately during propagation),
 }
  and the radius
of  {the} cophasal surfaces approaches infinity (i.e. $\tan^{-1}
(R)=\pi/2$).  {For the} special case  {where} $P_0=P_c$
 {the relation} $z_{c0}=z_R$  {is ensured and}
diffraction is exactly balanced by self-focusing. Therefore the beam
width and the radius of  {the} cophasal surfaces  {are} invariant
during propagation, and the breather  {reduces} to a soliton.

 \section{ Conclusion}
 In conclusion,  {optical beam} propagation in SNN media is connected with
 free propagation  {using} the technique of variable
transformation.
 The fact that the solutions as well as the
propagation properties in free space can be  {mapped to their
counterparts}  in SNN media through a one-to-one correspondence
makes  {this technique useful for investigating } the propagation
problems in SNN media. The efficient numerical method provided in
this letter  {is} of interest  {for} the investigation of beams with
irregular amplitude and phase profiles in SNN media.  {The technique
provides a} unified description of the existence conditions and
 {gives} possible structures of solitons and breathers in
SNN media.  {The various soliton and breather structures}
 {predicted herein} would greatly enrich the family of
solitons and breathers.

Mathematically, the modified SMM is equivalent to the famous
equation for the linear harmonic oscillator, which is widely used in
many branches of physics. Therefore, the relation proposed in this
letter can connect the free propagation not only with  { propagation
in SNN media}, but also with the evolution of waves in other systems
in which the governing equations  {reflect} the equation for the
linear harmonic oscillator. The  {technique described} in this
letter can be readily employed in other contexts with equivalent
governing equation, such as  {for} quadratic graded-index media
\cite{yariv},  {or for} harmonically trapped Bose-Einstein
condensates in the noninteracting limit \cite{bec1-li,bec2-Klein}.

 \section*{acknowledgements} This research was supported by the National
Natural Science Foundation of China (Grants No. 10674050 and No.
10804033), the Program for Innovative Research Team of Higher
Education in Guangdong (Grant No. 06CXTD005), and the Specialized
Research Fund for the Doctoral Program of Higher Education (Grants
No. 20060574006 and No. 200805740002).

\end{document}